# Signature of a silver phase percolation threshold in microscopically phase separated ternary $Ge_{0.15}Se_{0.85-x}Ag_x$ ($0 \leq x \leq 0.20$) glasses

Pulok Pattanayak and S. Asokan[a]
*Department of Instrumentation, Indian Institute of Science, Bangalore 560012, India*

Temperature modulated alternating differential scanning calorimetric studies show that Se rich $Ge_{0.15}Se_{0.85-x}Ag_x$ ($0 \leq x \leq 0.20$) glasses are microscopically phase separated, containing $Ag_2Se$ phases embedded in a $Ge_{0.15}Se_{0.85}$ backbone. With increasing silver concentration, $Ag_2Se$ phase percolates in the Ge–Se matrix, with a well-defined percolation threshold at $x=0.10$. A signature of this percolation transition is shown up in the thermal behavior, as the appearance of two exothermic crystallization peaks. Density, molar volume, and microhardness measurements, undertaken in the present study, also strongly support this view of percolation transition. The superionic conduction observed earlier in these glasses at higher silver proportions is likely to be connected with the silver phase percolation. © *2005 American Institute of Physics.* [DOI: 10.1063/1.1827341]

The interest in superionic glasses has increased considerably in the recent years because of the potential application of these materials in different fields.[1–3] Glasses on the selenium-rich side of the Ge–Se–Ag system are fast ionic conductors with silver as the mobile cation. Although several studies have been undertaken on Ag-doped chalcogenide glasses, especially on Ge–Se–Ag systems,[4–7] the microscopic structure of these glasses is not fully understood.

Ternary Ge–Se–Ag glasses can be formed in two distinct composition regions,[5,6,8] namely, a Se-rich region labeled as I, and a Ge-rich region labeled as II (Fig. 1). Se-rich glasses are found to be ionic whereas Ge-rich glasses are semiconductive. In this work, the thermal behavior of selenium rich Ge–Se–Ag glasses has been investigated using temperature modulated alternating differential scanning calorimetry (ADSC). Density, molar volume, and microhardness have also been measured. The results obtained have been found to throw light on the microscopic structure of these glasses.

Bulk Ge–Se–Ag glasses have been prepared by vacuum-sealed melt quenching method. Appropriate quantities of high purity (99.99%) constituent elements are sealed in an evacuated quartz ampoule (at $10^{-5}$ Torr) and slowly heated in a horizontal rotary furnace. The ampoules are maintained at 1000 °C and rotated continuously for about 24 h at 10 rpm to ensure homogeneity of the melt. The ampoules are subsequently quenched in a bath of ice water and NaOH mixture to get bulk glassy samples. The amorphous nature of the quenched samples is confirmed by x-ray diffraction.

Thermal analysis is carried out by ADSC (model DSC822$^c$, METTLER TOLEDO). ADSC scans of all samples are taken at a 5 °C min$^{-1}$ scan rate and 1 °C min$^{-1}$ modulation rate. Glass transition temperatures are deduced from the inflection point of the reversing heat flow curves and crystallization peaks from the nonreversing curves;[9] typical error in the measurements is within ±2 °C.

Figure 2(a) displays the reversible heat flow curves in the ADSC scan of $Ge_{0.15}Se_{0.85-x}Ag_x$ glasses. In all the curves, two endothermic glass transition peaks can be seen which show the heterogeneity of these glasses. Among these two, the lower glass transition peak $T_g^l$ systematically shifts up with the increase in Ag concentration. The higher glass transition peak $T_g^h$ does not shift much, but increases in strength with the increase in Ag concentration. It is seen in the present study that $\Delta C_p$ values of Ge–Se–Ag glasses, estimated from $T_g^h$, generally increases with Ag content and the increase becomes sharper at $Ag=0.10$ [Fig. 2(b)]. When the concentration of Ag is increased beyond 0.15, it is very difficult to distinguish the two glass transitions. If these samples are rerun in the ADSC, as suggested earlier,[5] only the first glass transition $T_g^l$ is observed as the second phase has already crystallized.

Figure 3(b) shows the nonreversible heat flow curves of $Ge_{0.15}Se_{0.85-x}Ag_x$ glasses in ADSC scans. One can discern two exothermic crystallization peaks ($T_{c1}$ and $T_{c2}$), at higher

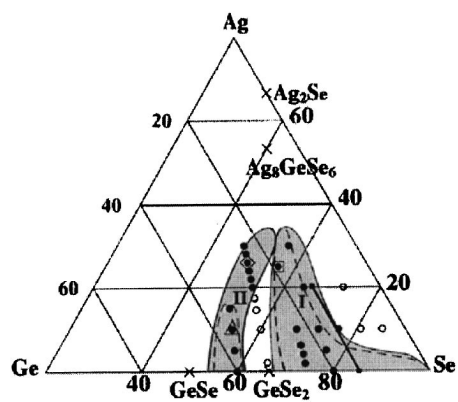

FIG. 1. The glass-forming regions in the Ge–Se–Ag ternary system (Ref. 6); the Se-rich and Ge-rich glasses are denoted by I and II, respectively. The solid black line in region I indicates samples chosen for the present work.

[a] Author to whom correspondence should be addressed; electronic mail: sasokan@isu.iisc.ernet.in

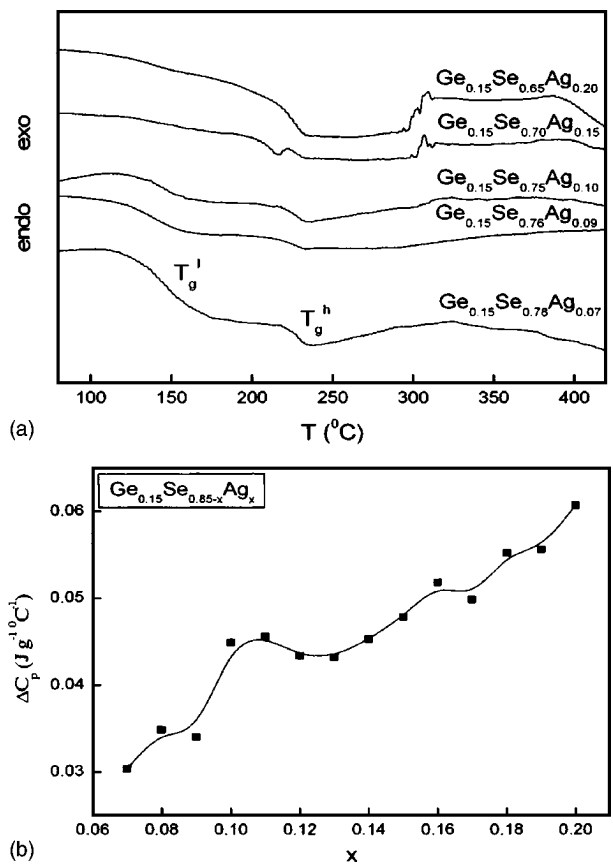

FIG. 2. (a) ADSC (reversible) curves showing the evolution of two glass transition temperatures $T_g^l$ and $T_g^h$ with increasing Ag concentration for representative $Ge_{0.15}Se_{0.85-x}Ag_x$ samples. (b) Variation of $\Delta C_p$ measured at $T_g^h$ with Ag concentration showing a sharp increase at Ag=0.10.

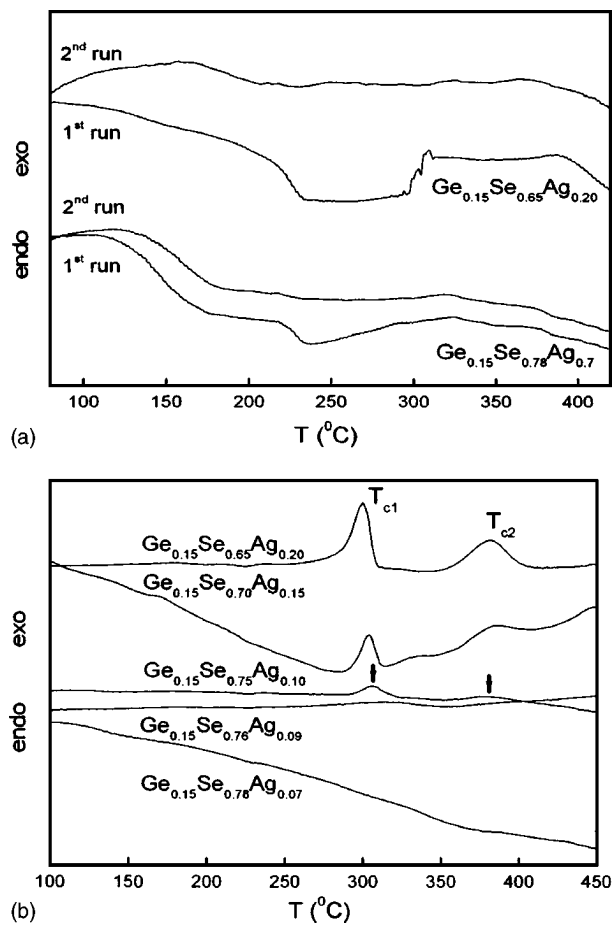

FIG. 3. (a) ADSC (reversible) curves of first and second runs for representative, $Ge_{0.15}Se_{0.78}Ag_{0.07}$ and $Ge_{0.15}Se_{0.65}Ag_{0.20}$ samples. (b) ADSC (nonreversible) curves showing the evolution of two crystallization exotherms ($T_{c1}$ and $T_{c2}$) with increasing Ag concentration for representative $Ge_{0.15}Se_{0.85-x}Ag_x$ samples. The first appearance of $T_{c1}$ and $T_{c2}$ marked by arrows is seen at $Ge_{0.15}Se_{0.75}Ag_{0.10}$.

Ag concentrations ($x \geq 0.10$), an important observation, which will be discussed later. It is also seen that the strength of both exothermic peaks increases systematically with Ag concentration afterwards.

Powder x-ray diffraction studies (using the Cu $K_\alpha$ line with $\lambda = 1.5405$ Å) of bulk glasses are taken after annealing at 320 and 450 °C (after the completion of crystallization reactions at $T_{c1}$ and $T_{c2}$, respectively) for several minutes to several hours to find out the products of the crystallization reactions. The x-ray reflections can be indexed for cubic $Ag_2Se$ and $c\text{-}Ag_8GeSe_6$ phases in samples annealed at 320 °C, but only $c\text{-}Ag_8GeSe_6$ is found in samples annealed at 450 °C (Fig. 4).

Density, molar volume, and microhardness studies are also carried out to find out the impact of the change in the microscopic structure of Ge–Se–Ag glasses. The density of the glasses is measured using a density measurement kit (model AG64, METTLER TOLEDO) with distilled water and $CH_3OH$ as reference liquids. The error in the density is estimated to be less than 2%. The Vickers microhardness of the samples is measured with an accuracy of ±5%, with a hardness tester (model HMV-2000, SHIMADZU). A pyramidal diamond indenter is used with an applied load of 50 g for 10 s. It is seen from the above measurements that a sharp transition is exhibited in the composition dependence of density, molar volume, and microhardness, around $x = 0.10$ [Figs. 5(a) and 5(b)].

Bimodal glass transitions of Se-rich Ge–Se–Ag glasses are attributed to Ge–Se backbone and $Ag_2Se$ inclusions.[5,6] As mentioned earlier, in the rerun of the samples, only one $T_g$ which corresponds to $Ge_{0.15}Se_{0.85}$ backbone is seen. However, it has been found that the glass transition temperatures in the second runs are marginally higher than those of the first runs [Fig. 3(a)]. Further, in the second runs, no crystallization reaction is observed (Fig. 6). This confirms that the two crystallization reactions observed are not related to $Ge_{0.15}Se_{0.8}$ backbones, contrary to the results obtained in Ref. 4, where $T_{c1}$ and $T_{c2}$ are attributed to $c\text{-}Ag_8GeSe_6$ and $c\text{-}GeSe_2$, respectively.

As mentioned above, $T_{c1}$ and $T_{c2}$ correspond to the crystallization of the $Ag_2Se$ and $Ag_8GeSe_6$ phases, respectively (Fig. 4). As $T_{c1}$ and $T_{c2}$ are close by, annealing at $T_{c1}$ shows mainly $c\text{-}Ag_8GeSe_6$ peaks along with cubic $Ag_2Se$ peaks which is formed as per the following reaction, suggested earlier:[5]

$$4(Ag_2Se) + GeSe_2 = Ag_8GeSe_6. \tag{1}$$

A simple crystallization scheme for the formation of the $Ag_8GeSe_6$ phase is as follows:

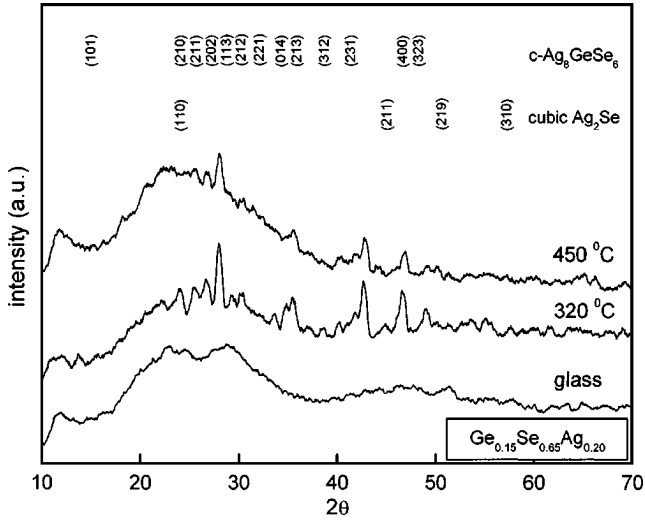

FIG. 4. XRD patterns of representative $Ge_{0.15}Se_{0.65}Ag_{0.20}$ samples as quenched and after different stages of annealing. The time of annealing for all the samples is 1 h.

$$Ge_ySe_{(1-y)-x}Ag_x = \left(\frac{x}{8}\right)Ag_8GeSe_6 + \left(1-\frac{x}{8}\right)Ge_tSe_{1-t}, \quad (2)$$

where

$$t = \frac{\left(\frac{y}{1-x}\right)\left(\frac{6x}{8}+1\right)-x}{1-\frac{x}{8}}. \quad (3)$$

In Eq. (2), the first term on the right-hand side designates the $Ag_2Se$ additive glass phase and the second term to the remaining base-glass phase.

Figure 7(a) shows that $T_g^l$ increases systemically with Ag concentration as the $Ge_{0.15}Se_{0.85}$ base glass becomes progressively Se deficient (and Ge-rich) as per Eq. (2), which is in agreement with the earlier result that the glass transition temperatures of $Ge_ySe_{1-y}$ glasses increases with Ge content, for compositions $y \leq 0.33$ (Ref. 10 and 11). Further, Mössbauer studies suggests that $Ge_ySe_{1-y}$ glasses are mainly comprised of $Ge(Se_{1/2})_4$ tetrahedral units for compositions $y < 0.33$ (Ref. 12). Recent NMR studies on $Ge_ySe_{1-y}$ glasses also reveal the presence of $Ge(Se_{1/2})_4$ tetrahedra and Se chains up to $y = 0.33$, without any Ge–Se–Se bonding.[13] As a result, inclusion of Ag in Se-rich $Ge_ySe_{1-y}$ glasses creates mainly $Ag_2Se$ phases. Based on the above, the phase-separated Ge–Se–Ag glasses can be described by[6]

$$Ge_ySe_{(1-y)-x}Ag_x = \left(\frac{x}{2}\right)Ag_2Se + \left(1-\frac{x}{2}\right)Ge_{t'}Se_{1-t'}, \quad (4)$$

where

$$t' = \frac{y}{\frac{x}{2}+y}. \quad (5)$$

While heating up to 450 °C, the $Ag_2Se$ phase crystallizes as $Ag_8GeSe_6$ following Eq. (1).

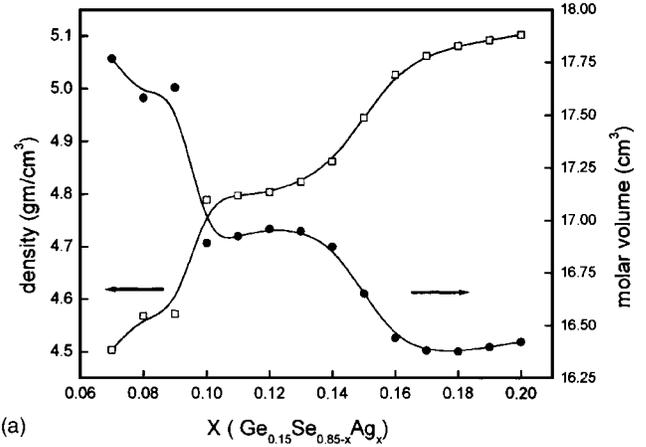

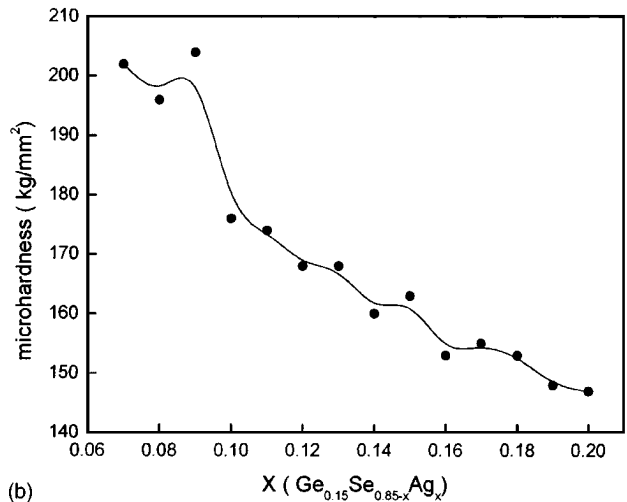

FIG. 5. (a) The variation of density and molar volume with Ag content. (b) The composition dependence of Vickers microhardness.

The most important observation in the present studies is the sudden appearance of the two crystallization exotherms ($T_{c1}$ and $T_{c2}$) at the composition $Ge_{0.15}Se_{0.75}Ag_{0.10}$ [Fig. 3(b)]. At this concentration, we also see a sudden jump in the $\Delta C_p$ variation for $T_g^h$ and not for $T_g^l$ [Fig. 2(b)]. This is a clear signature of the percolation threshold for the $Ag_2Se$ phase in these phase-separated glasses. At this concentration, the $Ag_2Se$ clusters embedded in $Ge_{0.15}Se_{0.85}$ base become homogeneous throughout the glass.

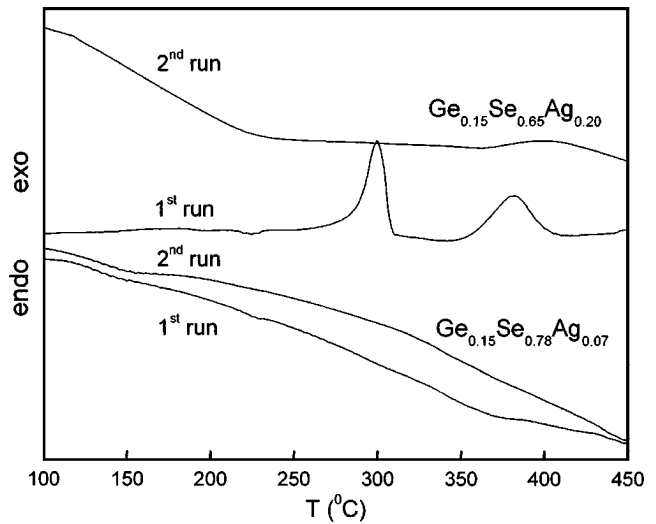

FIG. 6. ADSC (nonreversible) curves of the first run and second runs for representative $Ge_{0.15}Se_{0.78}Ag_{0.07}$ and $Ge_{0.15}Se_{0.65}Ag_{0.20}$ samples.

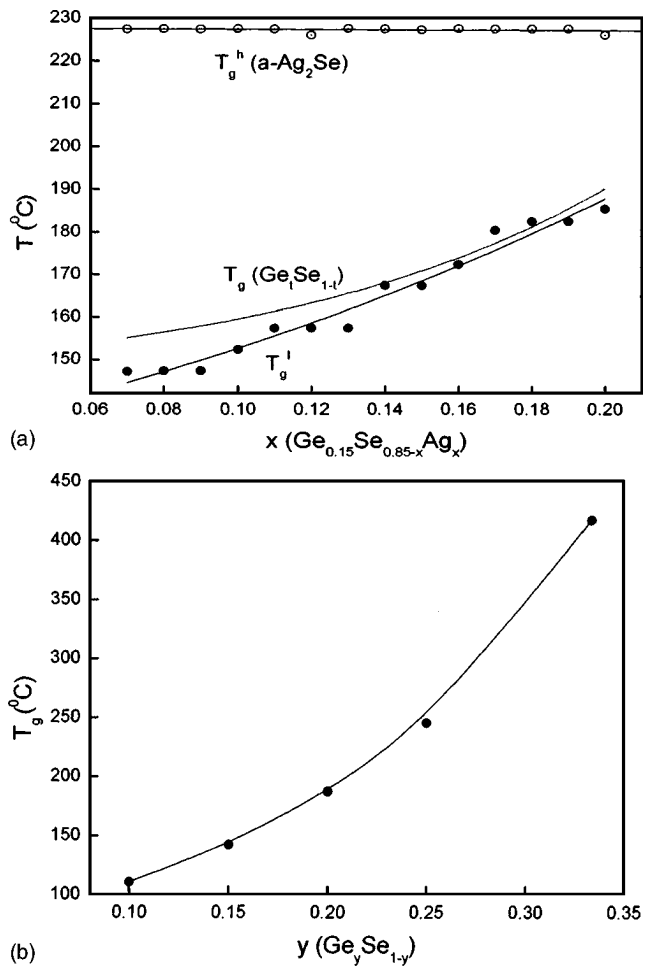

FIG. 7. (a) Evolution of bimodal glass transition temperatures $T_g^l$ and $T_g^h$ with Ag concentration for $Ge_{0.15}Se_{0.85-x}Ag_x$ glasses. The dotted curve gives the variation of $T_g^l$ following Eq. (2). (b) $T_g$ variation for $Ge_ySe_{1-y}$ glasses.

Studies on the variation of density, molar volume and microhardness of $Ge_{0.15}Se_{0.85-x}Ag_x$ glasses also show sharp transition at the Ag-phase percolation threshold [Figs. 5(a) and 5(b)]. As the $Ag_2Se$ inclusions in $Ge_{15}Se_{85}$ backbone become homogeneous, the density of total network increases sharply above x=0.10. For the same reason, a sudden reduction in the molar volume is found at this concentration. The appearance of $T_{c1}$ and $T_{c2}$ beyond Ag concentration of 0.10, clearly indicates the increase in the fragility of these glasses. Microhardness measurements are also in good agreement with this concept [Fig. 5(b)].

Percolation threshold of $Ag_2Se$ phase embedded in $Ge_{0.15}Se_{0.85}$ base is likely to be responsible for the sharp transition from semiconductive to ionic character of Se-rich Ge–Se–Ag glasses.[14,15] This reestablishes the fact that structural heterogeneity or granularity is the main criterion for ionicity in glasses.

Impact of silver phase percolation in electrical properties of these glasses is also exemplified in a sharp transition in the switching voltages at Ag=0.10, which will be discussed in more detail in forthcoming publications.[16]

In summary, the present study shows that $Ge_{0.15}Se_{0.85-x}Ag_x$ glasses are microscopically phase separated and composed of $Ag_2Se$ clusters and $GeSe_2$–Se network. When Ag concentration exceeds 10%, the $Ag_2Se$ clusters embedded in the $GeSe_2$–Se network percolates. The signature of this percolation threshold is clearly observed as the sudden appearance of two exothermic crystallization peaks in ADSC runs. Density, molar volume, and microhardness studies also strongly support this view of a percolation transition. The superionic conduction observed earlier in these glasses at higher silver proportions is likely to be connected with the silver phase percolation.